\documentclass[fleqn,usenatbib]{mnras}

\usepackage{newtxtext,newtxmath}
\usepackage{float}
\usepackage[T1]{fontenc}

\DeclareRobustCommand{\VAN}[3]{#2}
\let\VANthebibliography\thebibliography
\def\thebibliography{\DeclareRobustCommand{\VAN}[3]{##3}\VANthebibliography}

\usepackage{graphicx}	
\usepackage{amsmath}	

\title[$^{44}\mathrm{Ti}$ in thermonuclear supernovae]{Using $^{44}\mathrm{Ti}$ Emission to Differentiate Between Thermonuclear Supernova Progenitors}

\author[D. Kosakowski et al.]{
Daniel Kosakowski,$^{1}$\thanks{E-mail: dkosakowski@umassd.edu}
Mark Ivan Ugalino,$^{1}$
Robert Fisher,$^{1}$
Or Graur,$^{2,3}$
Alexey Bobrick,$^{4}$
and Hagai B. Perets$^{4}$
\\
$^{1}$Department of Physics, University of Massachusetts Dartmouth, 285 Old Westport Road, North Dartmouth, MA 02740, USA \\
$^{2}$Institute of Cosmology and Gravitation, University of Portsmouth, Portsmouth PO1 3FX, UK\\
$^{3}$Department of Astrophysics, American Museum of Natural History, New York, NY 10024, USA\\
$^{4}$Technion - Israel Institute of Technology, Physics Department, Haifa, Israel 32000
}

\date{Accepted XXX. Received YYY; in original form ZZZ}

\pubyear{2015}

\begin{document}
\label{firstpage}
\pagerange{\pageref{firstpage}--\pageref{lastpage}}
\maketitle

\begin{abstract}
The radiosotope $^{44}$Ti is produced through $\alpha$-rich freezeout and explosive helium burning in type Ia supernovae (SNe Ia). In this paper, we discuss how the detection of $^{44}$Ti, either through late-time light curves of SNe Ia, or directly via gamma rays, can uniquely constrain the origin of SNe Ia. In particular, building upon recent advances in the hydrodynamical simulation of helium-ignited double white dwarf binaries, we demonstrate that the detection of  $^{44}$Ti in a nearby SN Ia or in a young galactic supernova remnant (SNR) can discriminate between the double-detonation and double-degenerate channels of sub-Chandrasekhar (sub-$M_{\rm Ch}$) and  near-Chandrasekhar (near-$M_{\rm Ch}$) SNe Ia. In addition, we predict that the late-time light curves of calcium-rich transients are entirely dominated by $^{44}$Ti.
\end{abstract}




\section{Introduction} \label{sec:intro}

Thermonuclear supernovae play a fundamental role in astrophysics. For example, type Ia supernovae (SNe Ia) are standardizable candles used to measure extragalactic distances, which underpin the outstanding problems of dark energy and the Hubble tension \citep{riessetal1998, riessetal2021, perlmutteretal1999, Divalentinoetal2021, Freedman2021, Kenworthyetal2022}. Furthermore, SNe Ia are among the brightest explosions in the Universe and are thought to be the dominant source of Fe-peak elements in the Galaxy \citep{seitenzahletal13}. Similarly, Ca-rich supernovae, a recently discovered class of transients, inform us about detonation physics and progenitor properties of thermonuclear supernovae in general \citep{Zen+22}. Despite their astrophysical importance, the explosion mechanism and stellar progenitors of SNe Ia, and to a lesser extent of Ca-rich transients, remain unknown and subject to intense investigation \citep[see reviews by][]{ wangandhan2012, maozetal2014, soker2019}.  In particular, it is unknown whether normal SNe Ia occur primarily as a result of the thermonuclear explosion of a sub-$\rm{M}_{\rm{Ch}}$  or  near-$\rm{M}_{\rm{Ch}}$ carbon-oxygen white dwarf (C/O WD). It is therefore important to bring  additional observational probes to bear on the SN Ia progenitor problem.
\par

The most abundant radioisotope produced by Ca-rich transients and SNe Ia is $^{56}$Ni, which is primarily responsible for powering the early-time light curve \citep{pankey62}. However, because $^{56}$Ni is synthesized during nuclear statistical equilibrium conditions at high densities, the yield is insensitive over a wide range of C/O WD masses, and therefore a relatively weak probe of explosion channels \citep {diehletal2015}.
Two additional decay chains, produced by $^{57}$Ni and $^{55}$Co and their decay products, both the result of neutronized isotopes, have been detected in nearby SNe Ia and Ca-rich transients:

\begin {equation}
_{28}^{57}\textrm{Ni}\;\overset{\mathrm{1.5 d}}\longrightarrow\;_{27}^{57}\textrm{Co}\;\overset{\mathrm{272 d}}\longrightarrow\;_{26}^{57}\textrm{Fe}
\end {equation}

\begin {equation}
 _{27}^{55}\textrm{Co}\;\overset{\mathrm{18 h}}\longrightarrow\;_{26}^{55}\textrm{Fe}\;\overset{\mathrm{3 y}}\longrightarrow\;_{25}^{55}\textrm{Mn}
\end {equation}
These neutronized isotopes have been suggested as the powering mechanism behind the recently-discovered slow-down of SN Ia light curves at $\sim$ 800--2000 days past maximum light \citep{grauretal16, dimitriadisetal2017,kerzendorfetal2017,grauretal2018a,grauretal2018b,yangetal2018, lietal2019,graur2019,tuckeretal2021}. 
These decay chains signal nuclear burning with electron captures in high-density environments, and therefore characterize near-$M_{\rm Ch}$ WD progenitors. In contrast, due to their lower central densities, sub-M$_{\rm Ch}$ WD progenitors generally produce significantly less of these neutronized isotopes \citep{tiwarietal2022}. Motivated by these factors, studies have been performed seeking to use $^{57}$Ni and $^{55}$Co to distinguish between SNe Ia progentior scenarios \citep{ropkeetal2012}.
 To our knowledge, no such studies have attempted to use $^{44}$Ti, another product of SNe Ia. The neutronized isotopes are specifically informative for SN Ia spectra at times up to $\sim 2000$ d. For much later times, other radioisotopic signatures such as $^{44}$Ti (60 year half life) are required. 

Recent simulations of merging C/O white dwarfs with thin helium layers, leading to both full and surface-limited detonations, exhibit explosive helium burning, producing substantial amounts (10$^{-4}$ - 10$^{-3} \mathrm{M}_{\odot}$)  of the radioisotope $^{44}$Ti \citep {pakmoretal2022, royetal2022}. Because all but the most massive C/O WDs are predicted to have thin helium layers as the result of stellar evolution \citep {lawlormacdonald06}, this phase of explosive helium burning with its attendant nucleosynthesis of $^{44}$Ti will be generic to almost all WD mergers. Additionally, models of double-detonation SNe Ia predict similar yields $(\sim 10^{-3} \mathrm{M}_{\odot}$) of $^{44}$Ti \citep [e.g.] [] {leungnomoto20}. Finally, mergers with He-enriched WDs (He or hybrid HeCO WDs) can also give rise to enhanced production of $^{44}$Ti \cite{PeretsZenati20}.

In this paper, we discuss the use of $^{44}$Ti as a unique radioisotopic probe of sub-M$_{\rm Ch}$ WD progenitors. In \S 2, we present the relevance of $^{44}$Ti for SNe Ia. In \S 3 we examine the influence of $^{44}$Ti on the light curves of Ca-rich transients. In \S 4, we summarize existing direct gamma ray observations and show that $^{44}$Ti can be used to distinguish between different SN Ia classes.
Lastly, in \S 5, we summarize our findings and discuss some of their implications.

 %

\section{\texorpdfstring{$^{44}$}{}Ti Production in SNe Ia} \label{sec:relevance}


$^{44}$Ti is produced in both core-collapse supernovae (CC SNe) and SNe Ia. In CC SNe, it is most naturally produced in the freeze-out stage of nuclear burning. The density and temperature significantly affect the nuclear burning time scale, which determines whether a mass element undergoes normal freeze-out or alpha-rich freeze-out, with alpha-rich freeze-out being responsible for the production of substantial amounts of $^{44}$Ti 
\citep{woosleyetal1973, woosleyandhoffman1992, thielemannetal1990}. CC SNe inevitably undergo alpha-rich freeze-out and production of $^{44}$Ti \citep{woosleyetal1973, thielemannetal1996}. Because SNe Ia do not usually reach the conditions of alpha-rich freeze-out, an alternative mechanism is required to produce significant amounts of $^{44}$Ti, namely explosive helium burning \citep{woosleyetal1986,livneandarnett1995}.

\par A star must reach nuclear statistical equilibrium (NSE) in order for alpha-rich freeze-out to occur. During the NSE stage, alpha particles may merge into iron-group elements via the triple-alpha reaction on a timescale $\propto \rho^{-1/2}$, where $\rho$ is density \citep{timmesetal1996,magkotsiosetal2010}. 
A low density, rapidly cooling star would have a longer freeze-out timescale, leaving insufficient time for alpha particles to merge and consequently lead to the cooling down of nuclei in an alpha-rich environment \citep{timmesetal1996, blondinetal2021}. For SN Ia models that involve explosive helium burning, the helium layer of the primary never reaches NSE and subsequently does not undergo alpha-rich freeze-out \citep{timmesetal1996}. Explosive helium burning is characterized by the triple-alpha reaction rate and alpha capture reaction rate, with alpha capture likely to play a more significant role \citep{Khokhlov1984, shen2014initiation}. During this burning phase, $^{12}$C is produced via the reaction $3\alpha \longrightarrow ^{12}$C \citep{Khokhlov1984}. Further nucleosynthesis produces heavier elements up to $^{44}$Ti and beyond \citep{Khokhlov1984, khokhlovandergma1985}. If the helium is mixed with carbon, oxygen, or nitrogen, then alpha captures occur much more quickly than the triple-alpha reaction, resulting in a fast depletion of alpha particles \citep{shen2014initiation, gronow2020sne}. Consequently, nucleosynthesis of heavier elements such as $^{56}$Ni is suppressed and higher abundances of lighter elements such as $^{44}$Ti are produced.

\par
$^{44}$Ti has been directly measured in the CC SNR Cassiopeia A (Cas A) through gamma rays and SN 1987A through both gamma rays and its light curve. Additionally, $^{44}$Ti has been measured in SN 1987A by fitting the late-time light curve of the SN with the Bateman equation \citep{seitenzahletal2014}. Null detection of $^{44}$Ti lines in SNR of the peculiar SN 1885A give upper limits of $\sim0.12$ M$_{\odot}$ of $^{44}$Ti (W. Jianbin; private communication) which are still not constraining the models mentioned above. To date, the only type Ia SNR or SNe to have a measurement of $^{44}$Ti is the Tycho SNR \citep{trojaetal2014}. We return to the discussion of the direct detection of $^{44}$Ti in SNe Ia in gamma rays in \S4.  There is strong evidence that Ca-rich transients have large helium abundances and can produce substantial amounts of $^{44}$Ti \citep{perets2010, Zen+22}. Models predict these transients could produce as much as $\sim 10^{-1} M_{\odot}$ of $^{44}$Ti \citep{perets2010,waldman2011,Zen+22}. The production of $^{44}$Ti during explosive helium burning is therefore important for Ca-rich transients as well and can significantly impact the shape of their late-time light curves.

\section{\texorpdfstring{$^{44}$}{}Ti Powered Late-Time Light Curves of Thermonuclear Supernovae} \label{sec:lightcurves}
In this section, we calculate and plot the late-time pseudo-bolometric light curve of SN 2019ehk, a Ca-rich transient believed to originate from either a thermonuclear helium detonation or a low-mass stripped core-collapse event \citep{jacobsongalan2020, de2021,Zen+22}. We also make an approximation of the late-time light curve of SN 2011fe, an SN Ia, and consider whether it could be used to distinguish between a progenitor of the  sub-$\mathrm{M}_\mathrm{Ch}$ versus near-$\mathrm{M}_\mathrm{Ch}$ channels.

For the case of SN2019ehk, we calculate late-time pseudo-bolometric light curves obtained using a nebular model that self-consistently accounts for incomplete gamma-ray and positron trapping as the ejecta become optically thin \citep{cappellaro1997, jacobson2021}, 
\begin{equation}
    L(t) = S_{^{56}\mathrm{Co}}(t) + S_{^{44}\mathrm{Ti}}(t) 
\end{equation}
where contributions due to each isotope decay chain are denoted by a subscript. The contributions due to heating from radioactive decay of unstable isotope $^{44}$Ti 
is given by the Bateman equation solution \citep{seitenzahl2009late, Bateman1910}, 
\begin{equation}
    S_A(t) = M(A)\ \varepsilon_A\  e^{-\lambda_A t}
\end{equation}
where $\lambda_A$ is the decay rate, $M(A)$ is the mass of a radioactive isotope with atomic number $A$, and $\varepsilon_A$ is the specific average decay energy. 

We can manipulate the Bateman equation solution to obtain a condition that a more slowly-decaying radioisotope (e.g., $\lambda_B < \lambda_A$) always dominates the energy input of a more rapidly-decaying species,

\begin {equation}
{M (B) \over M (A)} > {\lambda_A \over \lambda_B} {\varepsilon_A \over \varepsilon_B}
\end {equation}
This expression yields the critical mass ratios M ($^{44}$Ti) /  M ($^{57}$Co) $\geq$ 115, and  M ($^{44}$Ti) /  M ($^{55}$Fe) $\geq$ 8.19, in order for $^{44}$Ti to dominate the late-time light curve entirely. The model of \citet{waldman2011} which most closely matches the kinetic energy and $^{56}$Ni yield of SN 2019ehk, the N14-enriched 0.2 $\mathrm{M}_\odot$ helium layer on top of a 0.5 $\mathrm{M}_\odot$ 
CO WD (model CO.5HE.2N.02), satisfies these nucleosynthetic criteria, strongly suggesting that the light curve of SN 2019ehk and similar Ca-rich transients will transition directly from a $^{56}$Ni-dominated phase to a $^{44}$Ti-dominated phase.

\begin{figure}
    \centering
    \includegraphics[width=3.5in]
    {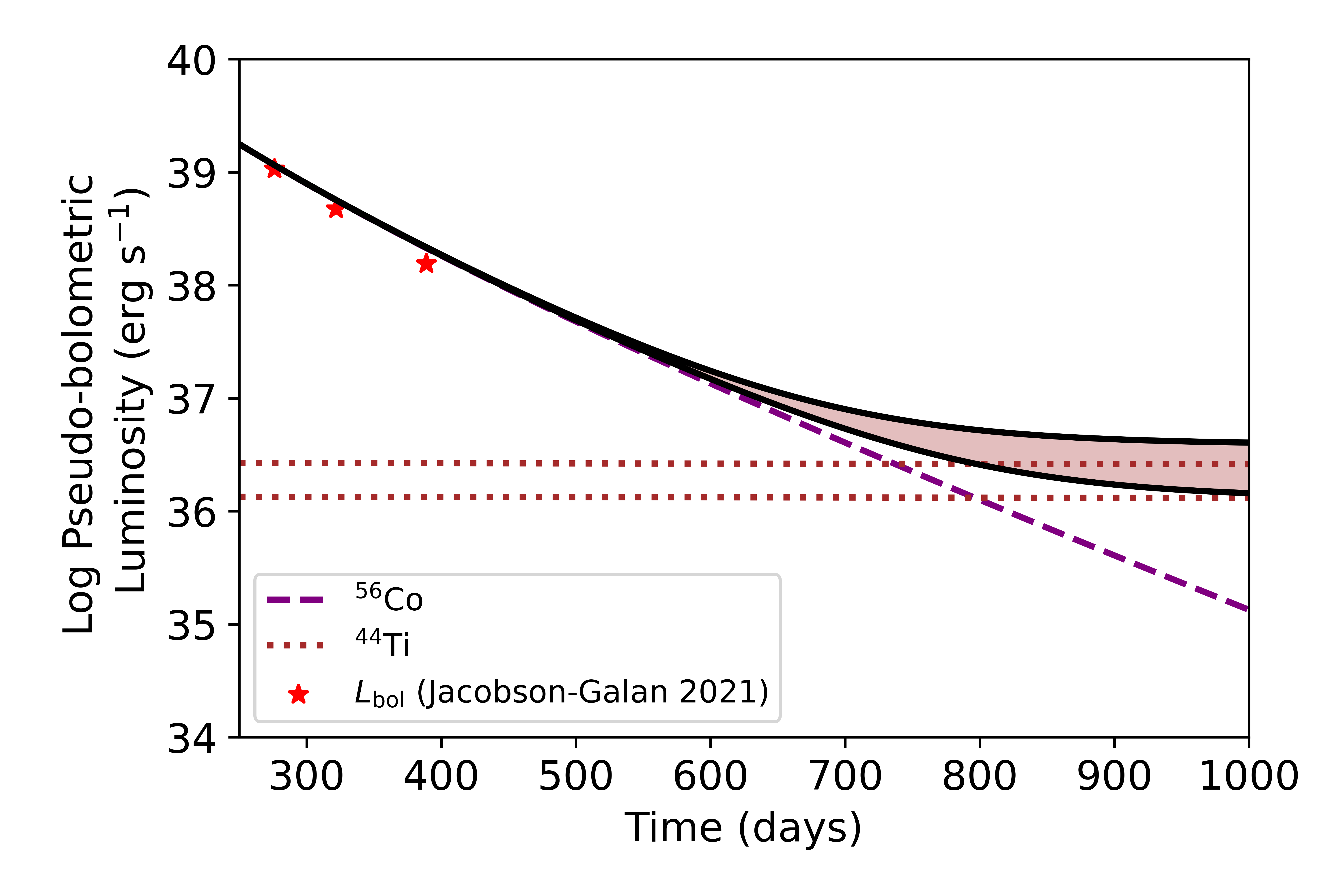}
    \caption{Late-time pseudo-bolometric light curves obtained for the Ca-rich SN 2019ehk using measured yields for $^{56}\mathrm{Co}$ from \citet{jacobson2021} ($[^{56}\mathrm{Co}] = 2.8  \times 10^{-2} \ \mathrm{M}_\odot$), and $^{44}$Ti from \citet{Zen+22}   ($[^{44}\mathrm{Ti}] = 7.34\times 10^{-3}-1.46 \times 10^{-2}\ \mathrm{M}_\odot$). The horizontal dotted lines represent the upper- and lower-bound contribution of $^{44}$Ti to the light curve.} 
    \label{fig:19ehk}
\end{figure}

Heating from gamma rays emitted from $^{56}$Co decay is given by,
\begin{equation}
    S_{^{56}\mathrm{Co}}^\gamma (t) = 0.81 \ S \ \left[1-
    e^{-(t_\gamma/t)^2}\right]
\end{equation}
while the energy released by gamma rays produced through positron annihilation corresponds to the following expression,
\begin{equation}
    S_{^{56}\mathrm{Co}}^+ (t) = 0.164\ S \ \left[1-
    e^{-(t_\gamma/t)^2}\right]\left[1-
    e^{-(t_+/t)^2}\right]
\end{equation}
where $S =M(^{56}\mathrm{Ni})\ \varepsilon_{^{56}\mathrm{Co}} \left[e^{-\lambda_{^{56}\mathrm{Co}}t}-e^{-\lambda_{^{56}\mathrm{Ni}}t} \right]$, and $t_+$, $t_\gamma$ are timescales for positron and gamma-ray trapping. We adopt the positron trapping timescale used in the late-time light curve fit by \citet{kerzendorfetal2017} $t_+ = 1200$ days, which is attributed to weak magnetic fields in the ejecta. We use the measured yields of $ [^{56} \mathrm{Co}] = 2.8 \times 10^{-2} \mathrm{M_{\odot}}$ from \citet{jacobson2021} and $[^{44}\mathrm{Ti}] = 7.34\times 10^{-3}-1.46 \times 10^{-2}\ \mathrm{M}_\odot$ from \citet{Zen+22} to obtain bounds for SN2019ehk (see Figure \ref{fig:19ehk}). We find that the light curve will be powered by $^{44}$Ti decay after $740-800$ days ($\sim 2.0-2.2$ yr) post-explosion.


\par

For the case of SN 2011fe, we use the Bateman solution and include the effects of $^{57}$Co and $^{55}$Fe. We take the estimated yields for $^{44}$Ti from simulated models of near-$\mathrm{M}_\mathrm{Ch}$ WD \citep{leung2018explosive}, sub-$\mathrm{M}_\mathrm{Ch}$ double degenerate WD mergers \citep{royetal2022,pakmoretal2022}, and sub-$\mathrm{M}_\mathrm{Ch}$ double detonations \citep{leungnomoto20}, generating lower and upper bounds for each channel. Using the expression above, the light curve from double degenerate mergers is estimated to be dominated by $^{44}$Ti decay past $(4.51-8.00)\times 10^3$ days ($\sim 12.3 - 21.9$ yr), and from double detonations $(4.01-9.71)\times 10^3$ days ($\sim 11.0-26.6$ yr) post-explosion. Lastly, for near-$\mathrm{M}_\mathrm{Ch}$ progenitors, $^{44}$Ti is projected to power the light curve after $ 9.3 \times 10^3 - 1.48 \times 10^4$ days ($\sim 25.5-40.5$ yr). This calculation is approximate, since it neglects the role of recombination and other atomic processes in the late-time light curve that could play a significant role. Detailed late-time light curve calculations including these atomic processes will be required to separate the $^{44}$Ti contribution from the recombination and atomic physics effects. However, the large separation of predicted timescales suggests the direct detection of $^{44}$Ti around one decade after the explosion would strongly point towards a sub-$M_{\rm Ch}$ explosion in  SN 2011fe.

\section{Direct Detection of \texorpdfstring{$^{44}$}{}Ti from Galactic SNe Ia in Gamma Rays} \label{sec:observe}

SNe may be observed through the gamma ray decay of radioisotopes synthesized in the explosion \citep{renaudetal2006}. Recent studies have found the upper gamma ray flux limit 
of two galactic SNRs, Kepler and G1.9+0.3, to be  $1.1\times 10^{-5}\text{ph cm}^{-2}\text{s}^{-1}$ and $1.0\times 10^{-5}\text{ph cm}^{-2}\text{s}^{-1}$, respectively, using a combined fit for the 68 keV, 78 keV $^{44}$Ti decay lines and the 1157 keV $^{44}$Sc decay line \citep{weinbergeretal2020}. The upper limit for Kepler corresponds to a mass limit of $4.0 \times 10^{-4}\mathrm{M}_{\odot}$, placing it just outside the range of $10^{-2}-10^{-3}\mathrm{M}_{\odot}$ predicted by previous double-detonation models \citep{weinbergeretal2020}. The Tycho Ia SNR has been detected with the Burst Alert Telescope(BAT)/Swift at a flux of $(1.3 \pm 0.6) \times 10^{-5}\text{ph cm}^{-2}\text{s}^{-1}$ for the 68 keV line and a flux of $(1.4 \pm 0.6) \times 10^{-5}\text{ph cm}^{-2}\text{s}^{-1}$ for the 78 keV line \citep{trojaetal2014}.

\par
\begin{figure}
    \centering
    \includegraphics[width=3.5in]{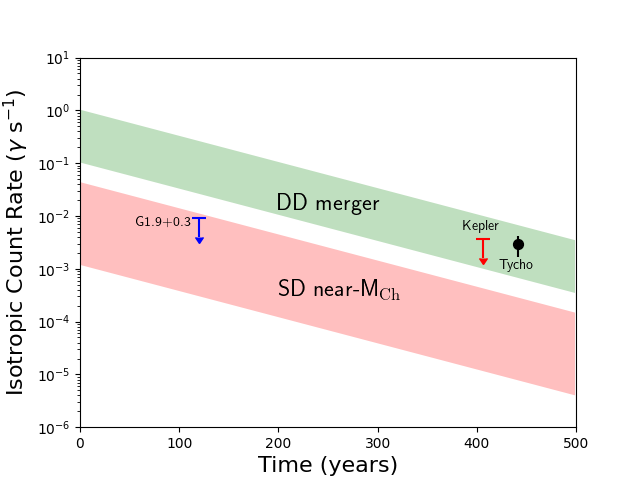} 
    \caption{Inferred isotropic photon count rate ($4\pi \times F_{\gamma}\times d_{kpc}^{2}$) for \textbf{double degenerate (DD) mergers}: $[^{44}\mathrm{Ti}] = 1.0 \times 10^{-4}\ -\ 1.0 \times 10^{-3}\ \mathrm{M}_\odot$ \citep{royetal2022, pakmoretal2022} and \textbf{single degenerate (SD) near-}$\textbf{M}_{\textbf{\rm{Ch}}}$: $[^{44}\mathrm{Ti}] = 1.15 \times 10^{-6}-4.25 \times 10^{-5}\ \mathrm{M}_\odot$ \citep{Leungnomoto2018} compared with Kepler, Tycho, and G1.9+0.3 SNRs. \textbf{Kepler}: $t = 406$\,yrs, $d = 5.1$\,kpc, upper bound $ F_{\gamma} = 1.1\times 10^{-5}\text{ph cm}^{-2}\text{s}^{-1}$\citep{weinbergeretal2020}. \textbf{Tycho}:  $t = 442$\,yrs, $d = 4.1$\,kpc, $F_{\gamma} = (1.4 \pm 0.6) \times 10^{-5}\text{ph cm}^{-2}\text{s}^{-1}$ at 78 keV line \citep{trojaetal2014}. \textbf{G1.9+0.3}: $t = 120$\,yrs, $d = 8.5$\,kpc, upper bound $F_{\gamma} = 1.0\times 10^{-5}\text{ph cm}^{-2}\text{s}^{-1}$\citep{weinbergeretal2020}.}
    \label{fig:intflux}
\end{figure}

 
The gamma ray line flux $F_{\gamma}$ is given by

\begin {equation}
F_{\gamma} = \frac{8.21 \times 10^{-3} M_{4} \exp(-t/87.7\ \rm {yr})}{d_{kpc}^{2}} \gamma\ \rm{cm}^{-2}\ \rm {s}^{-1},
\label {gammaflux}
\end {equation}
where $M_{4}$ is the mass of $^{44}$Ti in units of $10^{-4}\rm{M}_{\odot}$, $d_{kpc}$ is the distance in kpc from the Sun, and $\gamma$ denotes one photon \citep{theetal2006}.  We consider the distance-independent isotropic count rate, $4 \pi \times F_{\gamma} \times d_{kpc}^{2}$, as a function of age, and plot our results over 500 years (see Figure \ref{fig:intflux}), spanning the ages of the three SNRs we are interested in. For Tycho, we use the time of detection \citep{trojaetal2014}. For Kepler and G1.9+0.3, we use the time for which upper bounds were calculated \citep{weinbergeretal2020}. We consider the double degenerate merger 
channel, and the single degenerate near-$\rm{M}_{\rm{Ch}}$ channel. We adopt $^{44}$Ti mass ranges obtained from recent multidimensional hydrodynamical simulations;  $[^{44}\mathrm{Ti}] = 1.0 \times 10^{-4}\ -\ 1.0 \times 10^{-3}\ \mathrm{M}_\odot$ for He-ignited double degenerate mergers \citep{royetal2022, pakmoretal2022}, and $[^{44}\mathrm{Ti}] = 1.15 \times 10^{-6}\ -\ 4.25 \times 10^{-5}\ \mathrm{M}_\odot$  for  single degenerate near-$\rm{M}_{\rm{Ch}}$ SNe Ia \citep{Leungnomoto2018}. 

Significantly, the detection for Tycho is consistent with current numerical simulations of helium-ignited  double-degenerate mergers. In contrast, the upper bound for G1.9+0.3 is consistent with numerical simulations of the single degenerate near-$\rm{M}_{\rm{Ch}}$ event, but not with a double-degenerate merger. The upper bound for Kepler is consistent with models of either a double-degenerate event or a single degenerate near-$\rm{M}_{\rm{Ch}}$ event.  



\section{Conclusion} \label{sec:conclude}





In this paper, we have explored the use of  $^{44}$Ti as a unique probe of SN Ia progenitors. We discussed the production of $^{44}$Ti during explosive helium burning and how this implies that a substantial amount ($10^{-4}-10^{-3} M_{\odot}$) should be expected for almost any double-degenerate merger. Motivated by this, we considered the influence of $^{44}$Ti on the very late-time light curves of SN 2019ehk and SN 2011fe. We predict the light curve of SN 2019ehk will be entirely dominated by $^{44}$Ti after 550\ -\ 820 days. 
We also suggest that the light curve of SN 2011fe might become dominated by $^{44}$Ti decay at different time scales depending on whether it was a double-degenerate merger ( $(4.51-8.00)\times 10^{3}$ days), a double detonation ( $(4.01-9.71)\times 10^{3}$ days), or a single-degenerate near-$\rm{M}_{\rm{Ch}}$ explosion ( $9.3\times 10^{3}-1.48\times 10^{4}$ days). 
More detailed calculations incorporating recombination effects are required to make conclusive predictions about SN 2011fe. The final analysis carried out was a comparison of three Galactic SNRs with the predicted gamma ray flux of double-degenerate mergers and the single-degenerate near-$\rm{M}_{\rm{Ch}}$ channel. We find the Tycho detection to be consistent with a double-degenerate merger, while the upper bound for G1.9+0.3 is consistent with a single degenerate near-$\rm{M}_{\rm{Ch}}$ event and rules out a double-degenerate merger. Kepler's upper bound is consistent with either of the two channels.

There are two key caveats to the analysis carried out in this paper that need to be considered. One is that there is currently a limited parameter space for 3D hydrodynamical simulations of helium-ignited mergers. As future work is performed, it will be possible to put stronger constraints on the predicted ranges of $^{44}$Ti yields. The second caveat is the possibility of a significant amount of $^{44}$Ti being gravitationally bound and falling back onto the surviving remnant. The delayed decay of $^{44}$Ti could then drive winds from the surface of the surviving remnant and lead to a non-negligible contribution to the late-time light curve \citep{shenandschwab2017}. In fact this scenario can explain extreme winds in the galactic SN Iax candidate SN1181\citep{Lykouetal2022}. These two caveats require further work to be done which would strengthen the ability to use $^{44}$Ti as a means to distinguish between SN Ia progenitors. 

The production of $^{44}$Ti during explosive helium burning has implications not just for the SNe Ia progenitor problem, but also for the galactic positron problem. The 511 keV signal (due to positron annihilation) was first detected from the galactic center in the 1970's \citep{Johnsonetal1972}. After 50 years, the main source of these positrons remains uncertain. The decay chain  $^{44} \mathrm{Ti} \longrightarrow ^{44} \mathrm{Sc} \longrightarrow ^{44} \mathrm{Ca}$ produces these positrons, making the production of $^{44}$Ti a key ingredient \citep{Johnsonetal1972}. Since CC SNe are known to produce $\sim 10^{-4} \mathrm{M}_{\odot}$ of $^{44}$Ti, they have conventionally been thought to produce most of the $^{44}$Ti in the galaxy. However, the observed amount of galactic $^{44}$Ca is difficult to explain using only CC SNe \citep{theetal2006}. With SNe Ia capable of producing even larger amounts of $^{44}$Ti, they are increasingly considered to make a significant contribution to the galactic $^{44}$Ca abundance \citep{perets2014, crockeretal2017}. Recent arguments have shown that Ca-rich transients, such as SNR G306.3-0.9, alone are able to explain the 511 keV signal \citep{wengetal2022}. 

We note that current detectors such as BAT and INTEGRAL have line sensitivities $\sim 10^{-4} \text{ ph cm}^{-2}\text{s}^{-1} \text{and } 10^{-5} \text{ ph cm}^{-2}\text{s}^{-1} $, respectively \citep{skinneretal2008, savchenkoetal2017}. These current detectors are sufficient to constrain $^{44}$Ti yields of young SNRs within the Galaxy (Eqn. \ref {gammaflux} ).  Future planned telescopes such as COSI and AMEGO will reach sensitivities approaching $\sim 10^{-7} \text{ ph cm}^{-2}\text{s}^{-1}$ \citep{timmesetal2019}. Such high sensitivities will allow for direct gamma ray detection of  $\sim 10^{-4} M_{\odot}$ of $^{44}$Ti for any SN Ia remnant in the Galaxy, the Large Magellanic Cloud (LMC), Small Magellanic Cloud (SMC), and out to distances of up to $\sim 300$ kpc. These future detectors will also enable the full range of sub-$M_{\rm Ch}$ mass explosion progenitors to be confirmed or ruled out for Kepler and G1.9+0.3, and also begin to allow for meaningful constraints to be placed on the $^{44}$Ti abundances in the youngest SNRs in the LMC and SMC with distances of 50 - 60 kpc \citep {bozzettoetal17}. 
\par
Additionally, data from telescopes such as the James Webb Space Telescope (JWST), ELTs, Roman Space Telescope (Roman), and Athena will complement prospective gamma ray observations. Due to their sensitivity in the near-IR, JWST and Roman will make numerous observations of SNe Ia and their remnants. JWST will be able to to obtain spectra in addition to having the ability to detect and follow SNe at $z > 1$ \citep{gardneretal2010}. 
Roman is expected to observe $\sim 10^{4}$ SNe Ia, with the majority of these occurring within redshifts $0.5 < z  < 2$ \citep{joshietal2022}. Spectroscopic data will also be collected for some fraction of SNe Ia. Athena will be able to map ejecta abundance patterns and allow for the comparison of observed 3D SN Ia explosion properties with those of particular theoretical models \citep{barconsetal2012}. It will also be able to routinely measure elements such as $^{44}$Ti in young SNRs in the Galaxy, LMC, and SMC  \citep{barconsetal2012}. These advances will be transformative in either directly measuring or constraining the abundance of $^{44}$Ti in SNe Ia, and the associated mechanism of explosive helium burning on sub-$M_{\rm Ch}$ WDs.


\bibliography{ti44.bib}
\bibliographystyle{mnras}

\end{document}